\documentclass[times,doublespace]{article}

\usepackage{verbatim}
\usepackage{multirow}
\usepackage{amssymb} 
\usepackage{graphicx}
\usepackage{pifont,latexsym,ifthen,theorem,rotating,calc,textcase,booktabs,color}
\usepackage{amsfonts,amssymb,amsbsy,amsmath}
\usepackage[errorshow]{tracefnt}
\usepackage{fullpage}
\usepackage{setspace}

\usepackage{tikz}
\usetikzlibrary{mindmap,backgrounds, snakes, shapes, positioning}

\usepackage[colorlinks,bookmarksopen,bookmarksnumbered,citecolor=red,urlcolor=red]{hyperref}

\newcommand\startappendix{%
    \makeatletter 
       \setcounter{table}{0}
       \setcounter{figure}{0}
    \makeatother}

\newcommand\BibTeX{{\rmfamily B\kern-.05em \textsc{i\kern-.025em b}\kern-.08em
T\kern-.1667em\lower.7ex\hbox{E}\kern-.125emX}}

\begin{document}


\title{A Bayesian approach to the g-formula}

\author{Alexander P. Keil${}^{\text{\bf a}}$, Eric J. Daza${}^{\text{\bf b}}$, Stephanie M. Engel${}^{\text{\bf a}}$,\\ Jessie P. Buckley${}^{\text{\bf a}}$, Jessie K. Edwards${}^{\text{\bf a}}$}



\maketitle
${}^{\text{\bf a}}$Department of Epidemiology, University of North Carolina, Chapel Hill, NC, 27514 ${}^{\text{\bf b}}$ Stanford Prevention Research Center, Stanford University School of Medicine, Palo Alto, CA 94305

\section{Abstract}
Epidemiologists often wish to estimate quantities that are easy to communicate and correspond to the results of realistic public health scenarios. Methods from causal inference can answer these questions. We adopt the language of potential outcomes under Rubin's original Bayesian framework and show that the parametric g-formula is easily amenable to a Bayesian approach. We show that the frequentist properties of the Bayesian g-formula suggest it improves the accuracy of estimates of causal effects in small samples or when data may be sparse. We demonstrate our approach to estimate the effect of environmental tobacco smoke on body mass index z-scores among children aged 4-9 years who were enrolled in a longitudinal birth cohort in New York, USA. We give a general algorithm and supply SAS and Stan code that can be adopted to implement our computational approach in both time-fixed and longitudinal data.


\begin{onehalfspace}
\section{Introduction}

Epidemiologists often wish to estimate quantities that are easy to communicate and correspond to the results of realistic public health scenarios, such as interventions or policy changes. Recent developments in the field of causal inference have shown that, under some assumptions, observational data can be used to estimate such quantities \cite{Glass:2013ve}.  The parametric g-formula is one example of an approach from that field \cite{robins1986new}. Using the parametric g-formula, we can easily compare risks or rates of a health outcome in a population of interest under different exposure distributions \cite{Keil:2014jt}.

One assumption of the parametric g-formula is that we have accurately modeled the association between the outcome of interest and both the exposures of interest and potential confounders. For example, if the rate for a particular outcome varies in a linear-quadratic function with exposure, then our model could capture that by including both a linear and a quadratic term for exposure. This assumption can be satisfied by modeling the association using flexible models that incorporate non-linear functions and interaction terms. In particular realms, such as the study of complex exposure mixtures, complex longitudinal data, or small data sets, this may result in a model being "under identified", where we may be smoothing over strata in which we have little or no data, resulting in highly unstable estimates. Numerous statistical approaches have arisen for stabilizing model estimates when exposures come from a complex mixture or in other sparse data scenarios. Of note, Bayesian approaches have shown promise for dealing with under-identified models in environmental epidemiology through the use of parameter prior distributions to stabilize estimates that would be highly variable (or possibly inestimable) in an unpenalized maximum likelihood procedure \cite{maclehose2007bmh,Thomas:2007kx}. However, the issue remains that such approaches leave us with our original problem: an accurate model may be too complex to interpret usefully. Additionally in longitudinal settings, such standard approaches are generally not appropriate when exposure and confounders are time-varying \cite{robins1997causal}.


In the current manuscript, we demonstrate an approach that addresses the interpretability and the sparsity problems and extends to settings where exposures and confounders of interest vary over time. Building off prior work in Bayesian, causal inference  \cite{arjas2004causal,Wang:2011aa,saarela2015bayesian,Gustafson:2015aa} we outline a foundational Bayesian approach to the g-formula. In Section \ref{sec:bgf} we define notation and demonstrate the Bayesian g-formula in time-fixed data, and we discuss some potential settings of interest. We outline an algorithm in Section \ref{sec:bgfalgorithm} that can be used to develop software for our approach using existing Markov Chain Monte Carlo procedures. In Section \ref{sec:sims} we present simulations demonstrating our approach for estimating risk differences for simple exposure contrasts in two scenarios that are frequently problematic for epidemiologic inference: highly correlated exposure data and longitudinal studies with small sample sizes and time-varying confounders affected by prior exposure. In Section \ref{sec:application}, we demonstrate our approach using a longitudinal study of environmental tobacco smoke exposure and childhood body mass index (BMI). We close in Section \ref{sec:discussion} with a discussion of the assumptions underlying our approach and speculate about future extensions. In the Appendix, we derive our approach and we supply code in the Supporting Information to implement our approach for both the SAS and Stan programming languages.

\section{The Bayesian g-formula \label{sec:bgf}}
\subsection{The parametric g-formula}

To simplify our presentation, in this section we nest our example in the context of an observed set of time-fixed variables, $O = \big( Y, X, L \big)$, where $Y$ is a binary endpoint such as all-cause mortality, $X$ is a binary exposure of interest, such as the level (high or low) of criteria air pollutants, $\big( CO, NO_x, O_3, SO_2, Pb \big)$, and $L$ is a vector of potential confounders such as residence in an urban (versus rural) setting and annual income (above or below \$80,000, say). A more general setting is given in the Appendix.

In this setting, we are interested in the marginal distribution of the potential outcome $Y^g \in \{0, 1\}$, where $g$ refers to some intervention value (or distribution) for the exposures of interest. For example, we may be interested in the risk (i.e. cumulative incidence) of lung cancer in the population in which $g$ represents low concentrations of the pollutants. In other words, our intervention is to "set" each of the pollutants to "low." We assume that this hypothetical scenario corresponds to a well-defined intervention or set of interventions to reduce the pollutants to our specified level, such as mandating manufacturing changes to decrease vehicle exhaust emissions or restricting traffic on certain days. The risk under this intervention is given as $Pr\big(Y^g=1\big)$ \cite{Cole:2015aa}. For example, for binary $L$ the g-formula estimate of the risk under intervention can be expressed using the law of total probability as
\[
\Pr \left(
	Y^g = 1
\right)
=
	\sum_{ \ell \in \{ 0,1 \} }
		\Pr \left(
			Y^g = 1,
			L = \ell
		\right)
=
	\sum_{ \ell \in \{ 0,1 \} }
		\Pr \left(
			Y^g = 1
			\big|
			L = \ell
		\right)
		\Pr \left(
			L = \ell
		\right)
.
\]

To facilitate the expression of these quantities in a Bayesian framework, we adopt a more general notation and instead express the distribution of the potential outcome as
\begin{equation}
p \left(
	y^g
\right)
=
	\int_\ell
		p \left(
			y^g
			\big|
			\ell
		\right)
		p \left(
			\ell
		\right)
		d \ell
	\label{eqn:gfcontinuous}
,
\end{equation}
where the summation symbol has been replaced by an integral, and the discrete probability notation has been replaced by the generic probability function $p$. For a random variable $A$, if $A$ is discrete then we let $p ( a )$ denote $\Pr ( A = a )$, the mass of $A$ at realization $a$. Likewise, if $A$ is continuous then we let $p ( a )$ denote $f ( a )$, the density of $A$ at realization $a$. We also let $p ( \cdot | a )$ denote $p ( \cdot | A = a )$. To simplify notation we allow that the integral expression $\int g ( a ) d a$ for a function $g ( a )$ of $a$ will be used to denote integration if $A$ is continuous, i.e., $\int g ( a ) d a$, and summation if $A$ is discrete, i.e., $\sum_{ \{ a \} } g ( a )$. The formula given in \ref{eqn:gfcontinuous} yields the population distribution of the outcome under the intervention $g$, where the "population" referred to here is the set of individuals with the distribution of $L$ given by $p(l)$.

The parametric g-formula extends the g-formula given in \ref{eqn:gfcontinuous} by characterizing the conditional components $ p(y^g | \ell)$ and $p(\ell)$ using a parametric model. Thus, the parametric g-formula describing the data $O$ is denoted by 
\begin{equation}
p \left(
	y^g;
	\vphantom{\big|}
	\beta,
	\eta
\right)
=
	\int_\ell
		p \left(
			y^g
			\big|
			\ell;
			\beta
		\right)
		p \left(
			\ell;
			\eta
		\right)
		d \ell
=
	\int_\ell
		p \left(
			y
			\big|
			g,
			\ell;
			\beta
		\right)
		p \left(
			\ell;
			\eta
		\right)
		d \ell
	\label{eqn:parmgfcontinuous}
\end{equation}
where the parameter vector $\beta$ corresponds to the conditional change in the probability of $Y$ for unit changes in $X$ and $L$, and $\eta$ corresponds to the parameters of the probability of $L$. The rightmost side of \ref{eqn:parmgfcontinuous} is derived under the assumptions of consistency and conditional exchangeability, which are discussed further below. The distribution $p \big( y \big| g, \ell; \beta \big)$ could be modeled using logistic regression, and $\beta$ would correspond to the log-odds ratios. Summary effect measures, such as the causal risk difference for a unit increase in $X$ can be derived using Monte Carlo methods \cite{Keil:2014jt}. 

The utility of the g-formula becomes apparent when one considers that the functional form of the relationship between $X$ and $Y$ may be a complicated non-linear function in non-binary data, possibly with high order interaction terms. Note that, under such a scenario, the counterfactual distribution $p \big( \tilde{y}^g \big)$ is easily interpreted as the outcome distribution we would expect if we could have intervened to set $X$ to $g$. Similarly, if $L$ is high dimensional or $X$ (and possibly $L$) is time-varying, simple interpretations are still possible and can correspond to realistic settings. In other words, $\beta$ and $\eta$ are nuisance parameters - though their estimation is necessary for the parametric g-formula, our true interest is in $p \big( \tilde{y}^g \big)$. When $X$ and $L$ may affect each other over time, the g-formula can still be used to estimate unbiased net effects of exposure, whereas regression model estimates will generally be biased for this parameter \cite{rosenbaum1984consquences}.

In settings in which $L$ or $X$ is of very high dimension, such as when estimating effects in small samples, or when there is high correlation between elements of the data, the parameters $\beta$ or $\eta$ may not be stably estimated (i.e., due to variance inflation or finite sample bias). In such settings, common approaches are to accept some tradeoff in model fit or bias by creating a more parsimonious model, applying some penalization term, or adopting a semi or fully Bayesian framework to stabilize estimates using prior knowledge. We now adopt an approach that recasts the parametric g-formula within a Bayesian framework as a way to embrace the interpretability of the g-formula approach while employing the variance reduction properties of a Bayesian framework. That is, we allow the possible introduction of some bias in exchange for higher precision and an overall reduction in the mean squared error.

\subsection{A Bayesian approach}

Following Rubin \cite{rubin1978bayesian} and Saarela \cite{saarela2015bayesian}, we consider a Bayesian version of the potential outcome distribution, the posterior predictive distribution $p \big( \tilde{y}^g \big| o \big)$. The posterior predictive distribution of the potential outcomes given by the Bayesian approach differs from the estimated distribution of potential outcomes using the standard g-formula in that it marginalizes over the posterior distributions of the parameters ($\beta$ and $\eta$).


Our quantity of interest is the posterior predictive distribution of the potential outcome under the intervention $g$, In our simple example, this quantity is given as $p \big( \tilde{y}^g \big| o \big) = \int p \big( \tilde{y}^g \big| \theta \big) p \big( \theta \big| o \big) d \theta$ %
, where $p \big( \theta \big| o \big)$ is the posterior distribution of the parameters $\theta$ (which describe the models of the relationships between $Y$, $X$, and $L$). The distribution $p \big( \tilde{y}^g \big| o \big)$ takes as input the data and a value for $\theta$, and is used to generate a predicted outcome. For example, under consistency and conditional exchangeability, a predicted outcome can be generated using a generalized linear model for $Y$ by multiplying the design matrix for $X$ and $L$ by a draw from the distribution of the relevant parameters of $\theta$.

Drawing values from the posterior predictive distribution in our example data involves first drawing from the posterior distributions of $\beta$ and $\eta$. Next we draw from the distribution of the outcome given the "new" data $\big\{g, \tilde{L} \big\}$, where we "intervene" to set our exposure variable $X$ to the value $g$. Values of $\tilde{L}$ are drawn from the distribution of $L$ in the target population. If the distribution of $L$ in the study population (i.e., the empirical distribution) is thought to represent the distribution in the target population, then values of $\tilde{L}$ can be drawn from the empirical distribution of $L$. Mathematically, the posterior predictive distribution of the potential outcome under intervention is
\begin{align*}
p \left(
	\tilde{y}^g
	\big|
	o
\right)
& =
	\int_{ \tilde{\ell}}
		p \left(
			\tilde{y}
			\big|
			g,
			\tilde{\ell},
			o
		\right)
		p \left(
			\tilde{\ell}
			\big|
			o
		\right)
		d \tilde{\ell}
=
	\int_{ \tilde{\ell}}
	\int_{\theta}
		p \left(
			\tilde{y}
			\big|
			g,
			\tilde{\ell},
			\theta
		\right)
		p \left(
			\tilde{\ell}
			\big|
			\theta
		\right)
		p \left(
			\theta
			\big|
			o
		\right)
		d \theta
		d \tilde{\ell}.
\end{align*}
Under conditional exchangeability, consistency, and parameter independence assumptions detailed in the Appendix, this is a function only of the observed data, the target population distribution of $L$, the intervention value set by the analyst, and prior parameter distributions. We make these assumptions for the remainder of this section.

Before detailing the approach using the g-formula, it is worth noting that similar approaches have been implemented with other models that deal with counterfactual distributions $p \big( \tilde{y}^g \big)$, also known as structural models. Saarela et al.  \cite{saarela2015bayesian} recently outlined one approach using marginal structural models in which prior values can be put on parameters for the structural model (as well as the model for the inverse probability weights used to estimate the parameters of the model). Informative prior parameter values often come from conditional, non-structural outcome models, so the use of prior information in such settings is limited. As Robins et al point out, a Bayesian approach to estimating marginal structural model parameters with inverse probability weighting can also reduce some of the ability of the model to remove the covariate balance over exposure groups that results in confounding in finite samples \cite{Robins:2015aa}. 

In contrast, the Bayesian g-formula allows the incorporation of parameter prior distributions (i.e., "priors") often used by statisticians and epidemiologists. For example, one way of eliciting priors in a Bayesian analysis is to conduct a literature search and combine the parameter values found in this search through meta-analytic approaches. The Bayesian g-formula is amenable to these same priors because it relies on conditional regression models for the outcome, which are the usual starting point for epidemiologic inference. Additionally, the use of approaches such as hierarchical modeling and shrinkage estimation \cite{Greenland:2000gd} naturally fit the Bayesian g-formula framework, provided one can posit a reasonable set of hyper-parameters. 

For concreteness, suppose we have a binary outcome, exposure, and covariate, and a study population of $i = 1, \hdots, n$ independent observations. The Bayesian g-formula can be expressed in conventional notation involving the likelihood and the prior distribution. We define three parameters such that $\theta = \big\{ \beta, \alpha, \eta \big\}$ with likelihood function
\(
\mathcal{L} \left(
	\theta
	\big|
	o
\right)
=
	\mathcal{L} \left(
		\beta,
		\alpha,
		\eta
		\big|
		o
	\right)
.
\)
We also assume the parameters are independent of each other. The likelihood of the parameters given the observed data $O$ is the product of three binomial likelihoods:
\begin{align}
\mathcal{L} \left(
	\beta,
	\alpha,
	\eta
	\big|
	o
\right)
& =
	\prod_{i=1}^n
		p \left(
			y_i
			\big|
			x_i,
			\ell_i,
			\beta
		\right)
		p \left(
			x_i
			\big|
			\ell_i,
			\alpha
		\right)
		p \left(
			\ell_i
			\big|
			\eta
		\right)
	\label{eqn:likelihood}
\\
& =
	\prod_{i=1}^n
		\Pr \left(
			Y_i = 1
			\big|
			X_i = x_i,
			L_i = \ell_i,
			\beta
		\right)^{y_i}
		\Pr \left(
			Y_i = 0
			\big|
			X_i = x_i,
			L_i = \ell_i,
			\beta
		\right)^{1 - y_i}
		\times
	\nonumber
\\
& \quad \quad \; \;
		\Pr \left(
			X_i = 1
			\big|
			L_i = \ell_i,
			\alpha
		\right)^{x_i}
		\Pr \left(
			X_i = 0
			\big|
			L_i = \ell_i,
			\alpha
		\right)^{1 - x_i}
		\times
	\nonumber
\\
& \quad \quad \; \;
		\Pr \left(
			L_i = 1
			\big|
			\eta
		\right)^{\ell_i}
		\Pr \left(
			L_i = 0
			\big|
			\eta
		\right)^{1 - \ell_i}
	\nonumber
\end{align}
The posterior distribution of the parameters is given (up to a proportionality constant) using Bayes' formula, i.e.,
\begin{equation*}
p \left(
	\beta,
	\alpha,
	\eta
	\big|
	o
\right)
\propto
	\mathcal{L} \left(
		\beta,
		\alpha,
		\eta
		\big|
		o
	\right)
	\pi \left(
		\beta,
		\alpha,
		\eta
	\right)
,
\end{equation*}
where $\pi \big( \beta, \alpha, \eta \big)$ is the prior distribution of the parameters.

We note that \ref{eqn:likelihood} factors into three distinct binomial likelihoods, i.e.,
\(
\mathcal{L} \big(
	\beta,
	\alpha,
	\eta
	\big|
	o
\big)
=
	\mathcal{L} \big(
		\beta
		\big|
		y,
		x,
		\ell
	\big)
	\times
\)
\(	
	\mathcal{L} \big(
		\alpha
		\big|
		x,
		\ell
	\big)
	\times
	\mathcal{L} \big(
		\eta
		\big|
		\ell
	\big)
.
\)
This implies that $\beta$, $\alpha$, and $\eta$ can be estimated separately, for example, by fitting three different regression models. This also implies that we do not need to maximize $\mathcal{L} \big( \beta, \alpha, \eta \big| o \big)$ with respect to $\alpha$ in order to calculate the posterior estimates of $\beta$ and $\eta$. Under our assumptions, the potential outcomes under a static intervention do not depend on the probability of exposure \cite{Robins:2015aa}, as in the frequentist expression of the g-formula given in \ref{eqn:parmgfcontinuous}.

As we show in the Appendix, the posterior predictive distribution of $Y^g$ is proportional to:
\begin{align}
p \left(
	\tilde{y}^g
	\big|
	o
\right)
& \propto
	\int_{ \tilde{\ell}}
	\int_{ \eta}
	\int_{ \beta}
		p \left(
			\tilde{y}
			\big|
			g,
			\tilde{\ell},
			\beta
		\right)
		p \left(
			\tilde{\ell}
			\big|
			\eta
		\right)
		\mathcal{L} \left(
			\beta
			\big|
			y,
			x,
			\ell
		\right)
		\mathcal{L} \left(
			\eta
			\big|
			\ell
		\right)
		\pi \left(
			\beta
		\right)
		\pi \left(
			\eta
		\right)
		d \beta
		d \eta
		d \tilde{\ell}
	\label{eqn:bayesgf}
.
\end{align}

Using Markov Chain Monte Carlo (MCMC) methods, $p \big( \tilde{y}^g \big| o\big)$ is simply the distribution of $Y$ conditional on the data after setting the intervention value $g$ for $\tilde{X}$. Integration occurs by averaging the distribution across all posterior draws, after a suitable burn-in. 
%
In the time-fixed setting, or in time-varying settings when elements of $L$ occur temporally before any proposed intervention we can take a semi-Bayesian, semi-parametric approach to estimating \ref{eqn:bayesgf}. 
 Let $p_N \big( \ell \big)$ denote the empirical distribution of the observed covariates $\ell$.
This assumes that $\ell$ is a fixed set of values, and we set $\mathcal{L} \left(\eta\big| \ell\right) = p_N(\ell)$ and $\pi(\eta)=1$. Hence, $p \big( \eta \big| \ell \big) = 1$, and we set $\tilde{\ell} = \ell$. To estimate $p \big( \tilde{y}^g \big| o\big)$, models for $\pi \big( \beta \big)$ and $p \big( y \big| x, \ell, \beta \big)$ are specified, $g$ is set by the analyst, values of $\beta$ and $\tilde{y}$ are drawn, and the integration occurs by taking the sample average over all units. 

To conceptualize the computation, $\tilde Y^g$ acts like a missing variable that can be imputed after setting exposure \cite{Edwards:2015aa}. Thus, any MCMC software that can accommodate imputation of missing data may be used for the Bayesian g-formula. Computational considerations may necessitate finding closed-form solutions to \ref{eqn:bayesgf} when they are available and utilizing efficient sampling methods. For example, if there exists a closed-form solution for the full conditional distributions of each parameter, a Gibbs sampler may, in principal, be used. Frequently such solutions are not available, so other sampling schemes (e.g. rejection sampling) can be relied upon, but computational times will be longer. We describe in more detail an algorithm for this approach in section \ref{sec:bgfalgorithm}, which parallels an algorithm to estimate parameters of the frequentist g-formula \cite{Keil:2014jt}. This is also the approach that we implement in our simulations in section \ref{sec:sims} and real-world example in section \ref{sec:application}.

\section{Bayesian g-formula algorithm\label{sec:bgfalgorithm}}
The Bayesian g-formula is, by its fully Bayesian posterior formulation in \ref{eqn:bayesgf}, a single step procedure. Here we present an algorithm for estimation using the Bayesian g-formula that breaks that single step into multiple parts and corresponds to how one can numerically estimate quantities derived from \ref{eqn:bayesgf}, such as the difference in 30-year risk that compares the same group under two different exposures. In contrast with section \ref{sec:bgf}, in which all quantities were treated as time-fixed, we describe an algorithm for estimating this quantity using the Bayesian g-formula when exposure and covariates may be time-varying. We limit this approach to static, non-dynamic interventions such as "always treat" or "never treat," but our approach can be modified to include estimation of the outcome distribution under dynamic regimes such as "if at work, receive ergonomic training" or "if CD4 count drops below a threshold, initiate anti-retroviral treatment."

We expand on our time-fixed data example from the previous section by allowing for time-varying quantities. This requires choosing some time scale, such as time-on-study, calendar time, or age, and we denote specific points on that time scale as $t$. For simplicity, we treat $t$ as discrete. We denote the value of the exposure at time $t$ as $X(t)$ and let the history of exposure up to and including time $t$ be $\bar{X}(t) \equiv (X(0), ..., X(t))$. Time-varying covariate vectors $\boldmath{\bar{L}}(t)$ and outcomes ${\bar{Y}}(t)$ are denoted similarly. In our notation $\bar{X}(t)$, $\boldmath{\bar{L}}(t)$, and ${\bar{Y}}(t)$ are multivariate vectors with elements at times $0, ..., t$, but in practice these usually correspond to functional summaries over time that include only relevant aspects, such as the cumulative exposure through time $t$, possibly lagged for some period of time to account for latent disease. The vector $(\boldmath{L}(0), X(0), Y(0))$ contains the values of the covariates, exposure and outcome we observe at the start of the time scale of interest. Generally, $L(0)$ includes time-fixed quantities such as race, sex at birth, or income at baseline. We also denote the potential outcome of interest as $\tilde{Y}^g(t)$ and the posterior distribution of the covariate vector $L(t)$ under intervention $g$ as $\tilde{L}(t,g)$. These variable vectors are denoted in different ways to emphasize that we may be able to estimate the posterior distribution of the potential outcomes (a causal interpretation) under intervention $g$ without making the additional assumptions necessary to estimate the distribution of the other post-exposure covariates. We make the temporal assumption that $L(t)$ occurs temporally before $X(t)$, which implies that $\tilde{L}(0,g) = \tilde{L}(0)$, allowing us to non-parametrically sample $\tilde{L}(0)$ from the population empirical distribution $ p_N({l}(0))$, as noted in section \ref{sec:bgf}.

After we have selected an appropriate target population and intervention of interest, the Bayesian g-formula algorithm can be summarized in the following six steps:
%
%
%
%
\begin{description}
\item[1. Specify the model for the joint likelihood] Specify a joint model in the source population for $\{{L}(t), X(t), Y(t)\}$. For static regimes, this corresponds to the following
\begin{enumerate}

\item Specify a model for each component of $\boldmath{L}(t)$. These can be time-averaged, such as pooled logistic models that average coefficient values across time. If $\boldmath{L}(t)$ consists of a single component and $Y(t)$ is a death indicator (a counting process variable that equals 1 if the subject is dead at time $t$, 0 otherwise), this could be given as 
$$\mbox{logit}\left\{ L(t)=1| \bar{x}(t-1), \bar{l}(t-1), \bar{Y}(t-1)=\bar 0;\eta\right\} = \eta_0 + \eta_1t + \eta_2 \sum_{j=0}^{t-1} x(j) + \eta_3 \sum_{j=0}^{t-1} \ell(j)$$

\item Specify a model for $Y(t).$ Assuming again that $Y(t)$ is a death indicator, this model could be 
$$\mbox{logit}\left\{Y_1(t)=1| \bar{x}(t), \bar{l}(t), \bar{Y}(t-1)=\bar 0;\beta\right\} = \beta_0 + \beta_1t + \beta_2 \sum_{j=0}^{t} x(j) + \beta_3 \sum_{j=0}^{t} \ell(j)$$
\end{enumerate}
For presentation, we refer to these as separate models, though the Bayesian g-formula these are more accurately considered separate pieces of a single joint model for the data. Both of these relatively simple models should be considered for illustrative purposes only because they are likely over-simplifications for most realistic scenarios.

\item[2. Specify the prior distributions]
For a static intervention, we need only define prior distributions  $\pi(\beta)$ and $\pi(\eta)$ because modeling exposure is unnecessary. In most scenarios, investigators may be able to derive meaningful and relatively precise prior information for $\beta$ because this model often corresponds to regression models used in the literature. Parameters for $\eta$ may be more challenging because covariate model parameters are rarely reported. For both models, priors can be defined within a hierarchical modeling framework, such that the distributions of $\beta$ and $\eta$ may be determined by hyperprior distributions. In our applied example (\S \ref{sec:application}), we choose to use a hierarchical approach to stabilize the $\beta$ estimates. Other Bayesian solutions, such as the use of Dirichlet process priors, may also be a useful alternative in sparse data scenarios. 

In many scenarios, the analyst will be unable to specify informative, non-null distributions for some (or possibly all) parameters. For example, in most epidemiologic applications, $L$ is rarely of primary interest and is thus not modeled. However, one may specify reasonable generic information, such that the conditional log-odds ratios are likely to be within some interval or make use of weakly informative default priors that provide limited shrinkage but mostly let the likelihood drive the inference \cite{Greenland:2015aa}. In epidemiologic practice, it is rare to see (for example) odds ratios outside of the range 1/10 to 10, so even somewhat weakly informative priors that place such a range on the expected parameter values may justify the use of our Bayesian approach \cite{Hamra:2013aa}. 

\item[3. Sample from the target population] Define and sample from a target population defined by the distribution of $L(0)$. This is often identical to the source population, and $L(0)$ can be sampled by taking the empirical distribution of the baseline covariates in the data (the semi-Bayesian, semi-parametric approach). The target population will be considered to be under observation for some fixed time $k$, such that $t\equiv \{1,..,k\}$.

\item[4. Set the intervention values] Define a static intervention $\bar{g}(m)$ such as "always exposed", e.g., $g(t) = g = 1$ for all $t$. Set the exposures at all time points to this static intervention; i.e., $\bar{X}(m) \equiv g$. A parameter of interest could be the posterior predictive causal risk difference comparing "always exposed" to "never exposed."

\item[5. Draw from the posterior distribution of the model parameters] In our approach, draws from the posterior distribution of the model parameters given in Step 1 ($\hat\eta_0,...,\hat\eta_3,\hat\beta_0,...,\hat\beta_3$) are interpreted as posterior, conditional log-odds-ratios. These quantities are not of particular interest for our purposes, but are used in the next step to generate posterior predictive values. This is the point where typical Bayesian analyses stop, because the posterior log-odds ratio may be the primary parameter of interest in some settings. 

\item[6. Draw posterior predictive values] Draw new values of $\tilde{L}(t)$ and $ \tilde{Y}(t)$ in the target population by iteratively sampling values from the models in step 3 under intervention $g(t)$. Keeping our example, one could impute values of $ \tilde{L}_i(1)$ by simulating random values from a Bernoulli distribution with a mean equal to $logit^{-1}(\hat\eta_0 +\hat\eta_1t + \hat\eta_2  g(0) +\hat\eta_3 \ell(0))$. Similarly, $ \tilde{Y}(1)$ would be imputed as a draw from a Bernoulli distribution with a mean of $logit^{-1}(\hat\beta_0 + \hat\beta_1t + \hat\beta_2  g(0) + \hat\beta_3 (\tilde{\ell}(1,g) + \ell(0)))$. One would then similarly draw values of $ \tilde{L}(2)$ using $logit^{-1}(\hat\eta_0 + \hat\eta_1t + \hat\eta_2  (g(0)+g(1)) + \hat\eta_3 (\ell(0) + \tilde{\ell}(1))$ among those with $ \tilde{y}(1)=0.$ This process of sampling from conditional posterior distributions would continue iteratively for times $t=2,..., k$. Note that the posterior distributions of $\tilde L(t)$ and $\tilde Y(t)$ depend on the histories of the posterior distributions, denoted as  $\bar{\tilde L}(t-1)$ and $\bar{\tilde Y}(t-1)$, rather than the observed histories. By our assumptions, we have that $p\left({\tilde Y}(t)|g,o\right) = p\left({\tilde Y^g}(t)|o\right)$, such that draws from the posterior predictive distribution of $Y(t)$ (implicitly conditional on the data and the intervention) can be considered to be draws from the posterior predictive distribution of $Y^g(t)$.

\item[7. Estimate the parameter of interest] For the survival outcome $Y(t)$, we are interested in causal contrasts of the posterior potential outcome distribution $p(\bar{\tilde{y}}^g|o)$. For example, we can estimate the posterior potential risk difference at the end of follow-up, time $m$, defined as: $rd^{(1,0)}(m)\equiv E({\tilde{Y}}^1(m)) - E({\tilde{Y}}^0(m))$, by taking the difference of the sample posterior means of the imputed $\tilde{Y}_i^g(m)$ values:
$$
\widehat{rd}^{(1,0)}(m) = 
	n^{-1}\left(\sum_{i=1}^N\tilde{Y}_i^1(m)-\sum_{i=1}^N\tilde{Y}_i^0(k)\right) = 
	n^{-1}\sum_{i=1}^n\left(\tilde{Y}_i^1(m)-\tilde{Y}_i^0(m)\right).
$$
The 95\% posterior intervals could then be calculated by taking the 2.5th and 97.5th percentile of the posterior distribution of the risk difference or using an approximation such as taking $\pm 1.96$ times the standard deviation of the posterior distribution. 

We note that step 6 could be modified by considering only the expected values of $\{L(t,g), Y^g(t)\}$ at each time rather than simulating discrete values, as in Taubman et al's notable application of the frequentist g-formula \cite{Taubman:2009yq}. 
\end{description}

\section{Simulation study\label{sec:sims}}

We examine finite sample properties of a Bayesian g-formula approach using 2 simple simulation studies in scenarios in which the approach might be expected to perform reasonably well compared with the standard parametric g-formula. Our first simulation is in a time-fixed setting in which we have measured two exposures that are highly correlated with each other. Such settings are common in environmental epidemiology, such as in the air pollution literature where pollutants often come from common sources. Our second simulation deals with a simple, time-varying data scenario in small samples. Such settings are common in many longitudinal studies, where long follow-up on reasonably sized samples may lead to sparse strata.

For both simulations, we are primarily interested in ${rd}^{(\bar{1},\bar{0})}$, the causal risk difference comparing "always exposed" and "never exposed" (referent). The target population in both scenarios is the study population. We estimate $\widehat{rd}^{(\bar{1},\bar{0})}$ using both the frequentist g-formula ("standard g-formula") and the Bayesian g-formula. We calculate bias as $rd^{(\bar{1},\bar{0})}-{\widehat{rd}}^{(\bar{1},\bar{0})}$.  For each simulation, we perform the analyses on $M=1,000$ simulated datasets and average the bias over all samples.

For the standard g-formula, $\widehat{rd}^{(\bar{1},\bar{0})}$ and the standard error of the bias, were estimated using a non-parametric bootstrap. For many study designs, the bootstrap is the only feasible approach to variance estimation in the g-formula. For each of the $M$ iterations, we created $S=1,000$ bootstrap samples of each of the $M$ simulated datasets and calculated a risk difference for each by sampling with replacement from each dataset. The risk difference for each of the $M$ iterations was taken to be the mean risk difference across the $S$ samples, and the standard error was calculated as the sample standard deviation for these $S$ samples. Thus, for each analysis, the frequentist g-formula was repeated $S\times M = 1,000,000$ times.

For the Bayesian g-formula, we analyzed each dataset using MCMC methods implemented in the MCMC procedure of SAS 9.4. Each analysis was performed using $C=10,000$ iterations after a burn-in of $B = 1,000$ iterations. We estimated $\widehat{rd}^{(\bar{1},\bar{0})}$ using the sample mean of the risk difference across all MCMC iterations, and we calculated the standard error of  $\widehat{rd}^{(\bar{1},\bar{0})}$ by taking the sample standard deviation of the posterior estimates for the risk difference. Thus, for each analysis using the Bayesian g-formula, we made $(C+B)\times M = 11,000,000$ posterior draws.

We calculated Wald-type 95\% confidence intervals for each analysis by taking the estimated risk difference plus or minus 1.96 times the bootstrap/MCMC standard error. Confidence interval coverage was calculated as the proportion of all datasets for which the 95\% confidence intervals contained the true risk difference, $rd^{(\bar{1},\bar{0})}$. We also compared the standard deviation of the bias across the $M$ samples for each approach as a measure of precision. Mean squared error (MSE), our primary metric of comparison, was calculated as the average of the squared bias plus the variance over all samples. We compared the MSE between the Bayesian g-formula and the standard g-formula using the ratio of the MSE for the Bayesian g-formula and the MSE for the standard g-formula. A value of the MSE ratio below one indicates that the Bayesian g-formula performs better than the standard g-formula in a bias-variance tradeoff. 


We note that all simulations are performed on samples of 100 or fewer, primarily a choice driven by computational considerations stemming from our use of the non-parametric bootstrap and MCMC algorithms. Data generating mechanisms were designed to maximize precision of the parameter of interest in such small data situations, rather than to resemble any specific scenario. We expect that the choice between the Bayesian g-formula and the standard g-formula will be driven by the sparsity of data, which can also occur in large samples. In large samples sparse strata might occur when the outcome, exposure, or confounders of interest are rare in the study sample, or when two important predictors of the outcome are highly correlated. Thus, we expect that our simulation results will apply to larger sample sizes in more realistic scenarios.

\subsection{Time-fixed example\label{sec:tf1sim}}
\begin{figure}
\centering
\begin{tikzpicture}
\node(x){$X$};
\node(u)[above right=of x]{$U$};
\node(l)[left = of u]{$L$};
\node(y)[right = of x]{$Y$};
\path[-latex](u) edge (l);
\path[-latex](u) edge (y);
\path[-latex](l) edge (x);
\path[-latex](x) edge (y);

\end{tikzpicture}
\caption{Data generating mechanism for simulation given in \S \ref{sec:tf1sim} \label{fig:sim1dag}}
\end{figure}
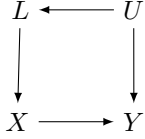

We simulate a simple dataset with two time-fixed exposures according to the graph shown in Figure \ref{fig:sim1dag}. The data are simulated to represent a scenario in which two highly correlated, measured variables may be independently associated with an outcome, and we are interested in the independent effect of only a single exposure. The simulated data consist of a binary outcome  $Y$, a binary exposure $X$ and a binary covariate $L$. The exposure of interest, $X$, is a direct cause of $Y$ while $L$ is an indirect cause of $Y$ through $X$. $L$ and $Y$ are also associated through a common cause $U$, which we consider to be unmeasured. $X$ and $L$ are simulated such that they have a known correlation coefficient, $\rho$.

The data are simulated according to the following data generating mechanism: 
\begin{itemize}
\item $U \sim uniform(0,1)$
\item Select $\nu_1-\nu_4$ (subject to the constraint $\sum_{i=1}^4\nu_i=1$) that define the correlation coefficient 
$\rho = {(\nu_1*\nu_4-\nu_2*\nu_3)}/{\sqrt{(\nu_1+\nu_3)*(\nu_2+\nu_4)*(\nu_1+\nu_2)*(\nu_3+\nu_4)}}$. These parameters correspond to the cell proportions in a cross-tabulation of $X$ and $L$. We additionally constrained the parameters such that $(\nu_1+\nu_2)=(\nu_1+\nu_3)=0.5$, i.e. both $X$ and $L$ had marginal expected values of 0.5.
\item $L = 1$ if $U<\nu_1+\nu_2$, 0 otherwise
\item Simulate $X$ according to the following scheme
\begin{itemize}
\item If $L=1$ then $X\sim Bernoulli(\nu_1+\nu4)$
\item If $L=0$ then $X\sim Bernoulli(\nu_2+\nu3)$
\end{itemize}
 \item $Y \sim Bernoulli(0.4+U/10 + rd^{(1,0)}\times X)$
\end{itemize}

We simulated data under a null effect ($rd^{(1,0)}=0$) and under  $rd^{(1,0)}=0.2$. We repeated each of these simulations for values of the correlation coefficient equal to (0.4, 0.8, 0.9, 0.98).

In a time-fixed setting, the standard g-formula requires only a single model to predict the outcome (as in the Bayesian g-formula, the standard g-formula can be semi-parametric if we take $p(\ell|\eta)$ to be the empirical distribution of $L$). For both the standard and the Bayesian g-formula, we predict $Y$ using a logistic model with terms for $X, \mbox{ and } L$. To assess the performance of the Bayesian g-formula even under low prior knowledge, we performed two sets of analyses for this simulation. In both settings, we used normal priors for all model coefficients. For the model intercept we used a vague prior with a normal distribution with a mean of log(0.5) and a variance of 1000 ($\mathcal{N}(log(0.5), 1000)$). All other coefficients were given prior distributions of $\mathcal{N}(0,3)$ (null, moderately informative priors). Each analysis was repeated under a true risk difference of $rd^{(\bar{1},\bar{0})}=0$ or $rd^{(\bar{1},\bar{0})}=0.2$. The "moderately informative" priors imply that we are 95\% certain that the true odds ratio corresponding to the coefficient lies between 1/30 and 30. 

Under moderately informative, null priors, the bias was larger for the Bayesian g-formula than for the standard g-formula (Table \ref{tab:sim1b}).  There was more variability in the standard g-formula estimator, however, as measured by the standard deviation of bias. With respect to MSE, the Bayesian g-formula under moderately informative priors outperformed the standard g-formula at correlation coefficients down to 0.8, suggesting that the utility of the Bayesian g-formula increases with exposure correlation. At a correlation of 0.4, the standard approach yielded a lower MSE, but we note that in that scenario, the MCMC standard error appeared to be slightly conservative (based on credible limit coverage higher than the nominal 95\%) for the Bayesian approach while the bootstrap standard error was slightly anti-conservative (based on the confidence limit coverage lower than the nominal 95\%).

\begin{table}
\centering{
\begin{tabular}{rccccccc}													
															\toprule
Method	&	Correlation($X,L$)	&	True RD	&	Mean bias	&	SD bias	&	MSE	&	Coverage	&	MSE ratio	\\\midrule
Standard	&	0.4	&	0	&	0.00	&	0.18	&	0.06	&	0.93	&	1	\\
Bayes	&	0.4	&	0	&	0.00	&	0.14	&	0.05	&	0.98	&	0.84	\\
Standard	&	0.4	&	0.2	&	0.00	&	0.11	&	0.02	&	0.94	&	1	\\
Bayes	&	0.4	&	0.2	&	-0.02	&	0.10	&	0.03	&	0.97	&	1.09	\\\\
Standard	&	0.8	&	0	&	-0.01	&	0.27	&	0.14	&	0.84	&	1	\\
Bayes	&	0.8	&	0	&	-0.01	&	0.15	&	0.07	&	0.99	&	0.48	\\
Standard	&	0.8	&	0.2	&	-0.01	&	0.17	&	0.06	&	0.92	&	1	\\
Bayes	&	0.8	&	0.2	&	-0.04	&	0.13	&	0.04	&	0.97	&	0.71	\\\\
Standard	&	0.9	&	0	&	0.00	&	0.29	&	0.16	&	0.81	&	1	\\
Bayes	&	0.9	&	0	&	-0.01	&	0.13	&	0.07	&	1.00	&	0.46	\\
Standard	&	0.9	&	0.2	&	-0.01	&	0.23	&	0.11	&	0.85	&	1	\\
Bayes	&	0.9	&	0.2	&	-0.06	&	0.14	&	0.06	&	0.97	&	0.52	\\\\
Standard	&	0.98	&	0	&	-0.03	&	0.23	&	0.10	&	0.90	&	1	\\
Bayes	&	0.98	&	0	&	-0.02	&	0.10	&	0.08	&	1.00	&	0.77	\\
Standard	&	0.98	&	0.2	&	-0.04	&	0.25	&	0.11	&	0.85	&	1	\\
Bayes	&	0.98	&	0.2	&	-0.11	&	0.11	&	0.08	&	1.00	&	0.68	\\\bottomrule
\end{tabular}															
}
\caption{Simulation scenario 1: Two correlated exposures, one time-fixed confounder model intercept priors were vague ($\mathcal{N}(ln(0.5),1000)$) and other coefficients were null centered and moderately informative ($\mathcal{N}(0,3)$)\label{tab:sim1b}}														
\end{table}
\clearpage
\subsection{Time-varying example\label{sec:tvcsim}}

\begin{figure}
\centering
\begin{tikzpicture}
\node(x1){$X(0)$};
\node(x2)[right = of x1]{$X(1)$};
\node(l)[above = of x2]{$L(1)$};
\node(y)[right = of x2]{$Y(1)$};
\node(u)[above=of y]{$U$};
\path[-latex](x1) edge (l);
\path[-latex](x1) edge (x2);
\path[-latex](u) edge (l);
\path[-latex](u) edge (y);
\path[-latex, in=-135, out=-45](x1) edge (y);
\path[-latex](x2) edge (y);
\path[-latex](l) edge (x2);

\end{tikzpicture}
\caption{Data generating mechanism for simulation given in \S \ref{sec:tvcsim} \label{fig:sim2dag}}
\end{figure}
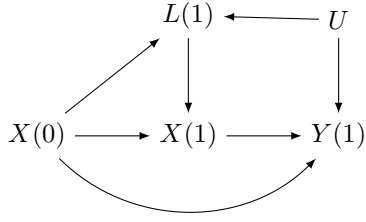

We simulate a simple dataset with time-varying exposure according to the graph shown in Figure \ref{fig:sim2dag}. The data consist of two time points, $t=(0,1)$, a binary outcome measured at the end of follow up $Y(1)$, binary exposure at two time points $\{X(0), X(1)\}$ a binary confounder of the exposure-outcome relationship that is also affected by prior exposure $L(1)$. $L(1)$ is related to the outcome by the common cause $U$, which we consider to be unmeasured. This particular simulation setup has been used previously to illustrate other causal inference methods \cite{robins2009estimation}, as it is feasible for addressing finite sample characteristics in computationally intensive approaches to time-varying data. It also addresses a central feature of the Bayesian g-formula: control of confounding by covariates that may also be affected by exposure. 

The data are simulated according to the following data generating mechanism:
\begin{itemize}
\item $U \sim uniform(0.4, 0.5)$
\item $X(0)\sim Bernoulli(0.5)$
\item $L(1)\sim Bernoulli(logit^{-1}(-1 + X(0) + U))$
\item $X(1) \sim Bernoulli(logit^{-1}(-1 + X(0) + L(1)))$
\item $Y  \sim Bernoulli(U + rd^{(\bar{1},\bar{0})}\times(X(0)+X(1))/2)$
\end{itemize}

In a time-varying setting, the parametric g-formula requires a model to predict each time-varying covariates and a model to predict the outcome. For both the standard and the Bayesian g-formula, we use the correct model (i.e. the logistic model implied by the data generating mechanism) to predict $L(1)$ and we predict $Y(1)$ using a logistic model with terms for $X(0), X(1), \mbox{ and } L(1)$. To assess the performance of the Bayesian g-formula even under low prior knowledge, we performed two sets of simulations. In both settings, we used normal priors for all model coefficients. We used null-centered priors for all model coefficients ($\mathcal{N}(ln(0.5),1000)$ intercepts and $\mathcal{N}(0,3)$ for other coefficients in models for $L$ and $Y$) . Each analysis was repeated under a true risk difference of $rd^{(\bar{1},\bar{0})}=0$ and $rd^{(\bar{1},\bar{0})}=0.2$. In the Supporting Information, we report on a sensitivity analysis for simulations in both time-fixed and time-varying scenarios in which we use priors with a larger variance (Tables \ref{tab:sim1a} and \ref{tab:sim2a}).

As in our time-fixed simulations, under moderately informative, null priors, the bias was larger and the variance was smaller for the Bayesian g-formula than for the standard g-formula.  The Bayesian g-formula under moderately informative priors outperformed the standard g-formula at sample sizes under 100 (with respect to MSE), suggesting that the utility of the Bayesian g-formula increases with reasonable, prior information (Table \ref{tab:sim2b}).

\begin{table}
\centering{

\begin{tabular}{rccccccc}															
															\toprule
Method	&	N	&	True RD	&	Mean bias	&	SD bias	&	MSE	&	Coverage	&	MSE ratio	\\\midrule
Standard	&	20	&	0.00	&	-0.01	&	0.35	&	0.26	&	0.90	&	1	\\
Bayes	&	20	&	0.00	&	0.00	&	0.23	&	0.14	&	0.97	&	0.55	\\
Standard	&	20	&	0.20	&	0.00	&	0.36	&	0.26	&	0.90	&	1	\\
Bayes	&	20	&	0.20	&	-0.06	&	0.23	&	0.14	&	0.98	&	0.56	\\\\
Standard	&	60	&	0.00	&	0.00	&	0.20	&	0.08	&	0.94	&	1	\\
Bayes	&	60	&	0.00	&	0.00	&	0.18	&	0.07	&	0.97	&	0.88	\\
Standard	&	60	&	0.20	&	-0.01	&	0.21	&	0.08	&	0.93	&	1	\\
Bayes	&	60	&	0.20	&	-0.03	&	0.18	&	0.07	&	0.96	&	0.88	\\\\
Standard	&	100	&	0.00	&	0.00	&	0.16	&	0.05	&	0.94	&	1	\\
Bayes	&	100	&	0.00	&	0.00	&	0.15	&	0.05	&	0.96	&	0.96	\\
Standard	&	100	&	0.20	&	0.00	&	0.15	&	0.05	&	0.95	&	1	\\
Bayes	&	100	&	0.20	&	-0.01	&	0.14	&	0.05	&	0.97	&	0.96	\\\bottomrule
\end{tabular}															
		}													
\caption{Simulation scenario 2: One exposure, one time-varying confounder that depends on exposure. Model intercept priors were vague ($\mathcal{N}(ln(0.5),1000)$) and other coefficients were null centered and moderately informative ($\mathcal{N}(0,3)$)\label{tab:sim2b}}															
\end{table}

\section{Application \label{sec:application}}

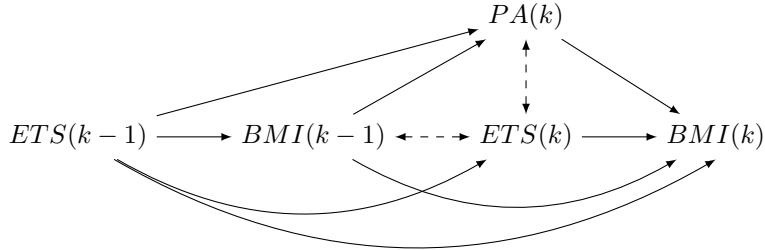
\begin{figure}
\centering
\begin{tikzpicture}
\node(x1){$ETS (k-1)$};
\node(y1)[right = of x1]{$BMI(k-1)$};
\node(x2)[right = of y1]{$ETS (k)$};
\node(y2)[right = of x2]{$BMI(k)$};
\node(l2)[above = of x2]{$PA(k)$};
\path[-latex, bend right](x1) edge (y2.south);
\path[-latex, bend right](x1) edge (x2);
\path[-latex](x1) edge (y1);
\path[-latex](x2) edge (y2);
\path[latex-latex, dashed](y1) edge (x2);
\path[-latex](y1) edge (l2);
\path[-latex, bend right](y1) edge (y2);
\path[-latex](l2) edge (y2);
\path[latex-latex, dashed](l2) edge (x2);
\path[-latex](x1) edge (l2);

\end{tikzpicture}
\caption{Proposed causal diagram for the example in section \S \ref{sec:application}. The effect of environmental tobacco smoke (ETS) at all times $1...k$ on body mass index (BMI) at time $k$ is our effect of interest, which may be confounded by physical activity (PA) and prior BMI by through unmeasured factors (dashed lines) that affect BMI, ETS, and PA. For simplicity, we ignore relationships between variables separated by more than one time point.  \label{fig:exdag}}
\end{figure}

We estimated the impact of interventions on environmental tobacco smoke on childhood BMI z-scores in a prospective birth cohort. The Mount Sinai Children's Environmental Health and Disease Prevention Research Center enrolled 479 primiparous women with singleton pregnancies in New York City between 1998 and 2002. The final cohort consists of 404 mother-infant pairs who met previously described inclusion criteria \cite{Engel:2007aa}. Children were invited to attend follow-up visits at approximately 4-5.5, 6-6.5, and 7-9 years of age (hereafter referred to as visits 1, 2, and 3, respectively) and 69 children attended all three visits. We assume that loss-to-follow-up is ignorable. Maternal baseline characteristics were ascertained via questionnaire at enrollment. At each follow-up visit, we calculated age- and sex-standardized BMI z-scores and classified children as physically active or inactive as described by Buckley et al \cite{Buckley:2015aa}.



We estimated posterior distributions for the BMI z-score using the Bayesian g-formula approach. Our specific approach involved a pooled-logistic model for physical activity and a pooled-linear model for the BMI z-score. The predictors in the physical activity model included all baseline covariates (maternal age [quadratic polynomial], maternal pre-pregnancy BMI [quadratic polynomial], maternal height [linear], smoking during pregnancy [yes, no], race [white, black, or other], education [some high school, high school grad, some college, college grad]) as well as time-varying covariates for cumulative number of visits at which physical activity was reported [lagged one visit, range 0-2], cumulative visits at which ETS was reported [lagged one visit, range 0-2], child's age (months), and product terms for cumulative ETS and all other variables. The model for BMI z-score used identical predictors and additional terms for current ETS, cumulative years of ETS, and current PA. For the physical activity model, we used moderately informative priors with $\mathcal{N}(0,3)$ distributions for all covariates, which represent a moderate skepticism that we captured any strong predictor of PA. For the BMI z-score model, we used a Bayesian LASSO (least absolute shrinkage and selection operator) approach similar to that of Park and Casella \cite{park2008bayesian}. Briefly, this approach utilizes a double exponential prior, assuming all regression coefficients are exchangeable. This approach differs from placing normally distributed priors on the coefficients in that it places much of the prior mass at the mean. We specified that the prior distribution for each parameter had a mean zero, again suggesting prior skepticism that any particular predictor of BMI z-score is strong. 

For visits 1, 2, and 3, we estimated the median BMI z-score under the interventions "always exposed to ETS" and "never exposed to ETS", as well as the difference in the mean BMI z-score under these interventions. We report the median of the posterior distribution of the population mean BMI z-score under each intervention (reported as "mean" or "Mean z-score"). For each intervention and difference in BMI z-score, we assessed statistical uncertainty using the 95\% credible intervals, which were estimated using the 2.5th and 97.5th percentile of the posterior distributions. 

Approximately 25\% of the children were exposed to ETS at visit 1, and this proportion increased to 35\% at visits 2 and 3. The average BMI observed among the children in the study was slightly above the age-specific population norms, with mean z-scores between 0.5 and 0.6 at visits 1,2, and 3 (Table \ref{tab:example}.) At all ages, we estimated that the BMI z-score was higher in the "always exposed" intervention group compared with the "never exposed" intervention group. Comparing the BMI z-score in the always exposed to never exposed (referent), the difference in the mean BMI z-score was largest at visit 2 (posterior median [95\% CrI] = 0.42 [-0.22, 1.05]), though the credible intervals included the null at all ages. Our results suggest that environmental tobacco smoke is associated with higher BMI z-score throughout childhood, though the small study size precludes precise estimation in the study population. 

\begin{table}[htdp]
\caption{Comparing childhood BMI z-scores after potential interventions on environmental tobacco smoke exposure in 69 children, New York, USA 2004 to 2008.}
\begin{center}
\begin{tabular}{r cl  cl   cl  cl  }\toprule	& \multicolumn{2}{c}{Observed}	& \multicolumn{6}{c}{Intervention}		\\\cline{4-9}								
&&&	\multicolumn{2}{c}{Never exposed}&\multicolumn{2}{c}{Always exposed} & \multicolumn{2}{c}{Difference}	\\
	\multicolumn{1}{c}{Visit} & \multirow{2}{.5in}{ Mean  z-score} && \multirow{2}{.5in}{ Mean z-score} && \multirow{2}{.5in}{ Mean z-score}&&\multirow{2}{.5in}{ \\\hfill Mean } \\
\multicolumn{1}{c}{(age$^a$)} &  & \hfil 95\% CI&& \hfil95\% CrI && \hfil95\% CrI && \hfil95\% CrI  \\\midrule
1 (5)	&	0.52	&	(0.22, 0.81)&	0.35	&	(0.13, 0.57)	&	0.64	&	(0.27, 1.02)	&	0.29	&	(-0.14, 0.73)	\\
2 (6)	&	0.55	&	(0.29, 0.80)&	0.46	&	(0.19, 0.73)	&	0.88	&	(0.32, 1.43)	&	0.42	&	(-0.22, 1.05)	\\
3 (7)	&	0.59	&	(0.32, 0.85)&	0.44	&	(0.10, 0.79)	&	0.80	&	(0.07, 1.54)	&	0.36	&	(-0.49, 1.21)	\\\bottomrule
\end{tabular}																	\end{center}
\label{tab:example}
$^a$ Mean age at visit
\end{table}%

While our results should be seen as primarily illustrative, they show that the effects of interventions on childhood risk factors for obesity can be estimated using epidemiologic data. Such inference is valuable in placing the results of observational studies in real-world contexts to prioritize public health actions. We could easily compare the benefits of interventions on smoking to the benefits of interventions on physical activity or diet, for example. Such comparisons are difficult with standard regression approaches, even under conditions in which estimators from regression models are unbiased. 

One potential weakness of the LASSO approach is that we assume that all model coefficients are exchangeable. It is likely that prior BMI is a much stronger predictor of current BMI than the other confounders considered, so future applications of the Bayesian g-formula could implement non-parametric Bayesian approaches that allow coefficients to shrink to different means. To assess whether our results might be sensitive to the shape of the LASSO prior, we estimated the risk difference using null-centered, normal variables ($\mathcal{N}(0,3)$) in the model for the BMI z-score. The new priors resulted in changes to the risk difference at each visit of less than 0.03 (7\%), and results were slightly more precise. Thus, even with the limitations of our illustrative approach, our results appear to be robust to a reasonable range of priors.

\section{Discussion\label{sec:discussion}}
From a Bayesian perspective, the g-formula is a natural approach causal inference. The frequentist version of the g-formula utilizes the basic rules of probability calculus, along with identification assumptions, in order to make inference on the effects of treatment regimes or interventions. The Bayesian g-formula adds the prior probability distributions into the mathematical framework of the g-formula - the novelty lies in which posterior inferences are of interest. Indeed, our simple time-fixed simulation corresponds to the same use of Bayesian inference that typifies epidemiology -- posterior joint distributions of the exposures and outcomes of interest --  but we allow that the probability distribution of the data is also informative regarding intervention effects. We argue, however, that aside from the numerous philosophical reasons for Bayesian causal inference, the reduction of the MSE makes it attractive for data analysis from a frequentist perspective. Our simulation results suggest that we can improve accuracy over standard g-formula analyses even with very limited prior information, especially when effects are under-identified due to high correlation among exposures or in small samples. Our example in \S \ref{sec:application} shows that the Bayesian g-formula is useful in the epidemiologic setting that originally motivated the g-formula: longitudinal design in which exposures, covariates, and outcomes may change over time \cite{Keil:2014jt}. The approach is useful for making easily interpretable inference, even when data or the relationships between variables are highly complex.

Some of the usual assumptions made in causal inference must be slightly modified under our approach. Namely, we must make Bayesian versions of the assumptions of exchangeability, and positivity, which we describe further in the Appendix. Notably, these assumptions do not differ greatly from their frequentist counterparts. In the Bayesian setting, we may be less prone to bias due to random deviations from positivity, because our priors help to smooth over the empty strata. We make the usual assumptions regarding counterfactual consistency\cite{Pearl:2010bh} and non-interference\cite{Hudgens:2008mi}. We also assume correct specification of the likelihood. In the case in which the likelihood is correctly specified (or approximately so), the use of priors may lead to biased estimates, as we showed in our simulations. However, the use of shrinkage-type priors leads to an overall improvement in accuracy in the frequentist sense, in terms of MSE reduction. An issue that is not resolved by moving to a Bayesian framework is the g-null paradox, in which model misspecification is induced by including an exposure that has no causal effect on the outcome \cite{robins2003general}. This paradox leads to hypothesis tests that inevitably reject the null hypothesis as sample size increases. The implications of this paradox for finite samples are unknown, however.

Our approach has much in common with recent examples of Bayesian causal inference. Namely, Arjas and co-authors developed an approach analogous to the Bayesian g-formula under a causal framework that does not utilize potential outcomes  (described in a general setting in \cite{arjas2004causal} and under a Bayesian framework in chapter 7 of \cite{berzuini2012causality}, with a practical example given in \cite{arjas2010optimal} and \cite{Saarela:2015ab}). Robins calls this the 'agnostic' causal model \cite{robins2000causal}. In contrast, our approach is derived under the original causal framework of Rubin, which bases causal inference on the concept of potential outcomes \cite{rubin1978bayesian}. A previous example by Wang et al. used the Bayesian g-formula to estimate the effect of a change in ventilation practices on survival in patients with acute lung injury \cite{Wang:2011aa}. Unlike those authors, who used vague normal priors, we utilized a hierarchical shrinkage prior, demonstrating the facility with which other types of priors can be incorporated. Our manuscript builds on the work of Wang et al. by improving the mathematical foundations underlying the Bayesian g-formula. Further, we provide a practical example along with SAS and Stan code that can be used to carry out our simple simulation using the g-formula for a time-varying exposure using MCMC methods. Because standard g-formula algorithms typically require large Monte Carlo samples and a non-parametric bootstrap, the use of MCMC machinery in the Bayesian g-formula does not result in a substantial computational burden relative to the frequentist approach.

In the current manuscript, we frame our questions about population health in terms of interventions. In this way, the Bayesian g-formula is useful for estimating parameters with direct public health implications. Because our approach quantifies the effects of interventions on exposure distributions, it is useful to inform policy. In environmental studies, such as air pollution studies, we are often confronted with highly correlated exposures that change over time. Using our approach, we can directly estimate the public health impact we could expect under interventions that changed the distribution of pollutants in the air. Using Bayesian decision analysis and  quantifying both the costs of interventions and the utility of health outcomes, we could compare interventions in terms of cost effectiveness. The recent availability of powerful personal computers and the rapid development of probabilistic programming languages are increasingly allowing for Bayesian inference to be possible in realms in which the computational demands were previously too high. By bridging the worlds of Bayesian inference and causal inference, we can unleash this very unique and powerful set of tools to answer challenging public health questions.



%

\end{onehalfspace}
\pagebreak


\bibliographystyle{unsrt}

\startappendix
\section{Appendix - Derivations}
 \setcounter{table}{0}
\setcounter{figure}{0}
\setcounter{equation}{0}
\renewcommand\theequation{A\arabic{equation}}
\renewcommand\thetable{A\arabic{table}}

Here we derive a Bayesian approach to the g-formula. We first review the g-formula as a way to estimate counterfactual distributions in the frequentist setting, and discuss how it is parameterized. We then describe the Bayesian analogue to counterfactuals and describe how their distributions can be estimated in a Bayesian framework.

\subsection{Notation}
For any random variable $A$, if $A$ is discrete then let $p ( a )$ denote $\Pr ( A = a )$, the mass of $A$ at $a$. Likewise, if $A$ is continuous then let $p ( a )$ denote $f ( a )$, the density of $A$ at $a$. Let $p ( \cdot | a )$ denote $p ( \cdot | A = a )$, and likewise let $E ( \cdot | a )$ denote $E ( \cdot | A = a )$, where $E ( \cdot )$ denotes the expectation function. In a slight abuse of notation, the integral expression $\int g ( a ) d a$ for a function $g ( a )$ of $a$ will be used to denote integration if $A$ is continuous, i.e., $\int g ( a ) d a$, and summation if $A$ is discrete, i.e., $\sum_{ \{ a \} } g ( a )$.Suppose we have an observed sample of $i = 1, \hdots, n$ measurements. Let $Y$, $X$, and $L$ denote random variables or vectors that represent an outcome, a binary exposure or treatment such that $X = 0, 1$, and other covariates, respectively. Let $O = \big\{ Y, X, L \big\}$ represent these observed quantities, where $A = \big\{ A_1, \hdots, A_n \big\}$ for a random variable $A = Y, X, L$. Let $Y^g$ represent the potential outcome $Y$ under exposure or treatment $X=g$.

\subsection{Frequentist Approach}Suppose that the following assumptions are true. 
\begin{itemize}
	\item Conditional exchangeability: $Y^g \perp\!\!\!\perp X \big| L$ for $x= 0, 1$ 
	\item Consistency: $Y = Y^1 X + Y^0 ( 1 - X )$ where the support of $Y^1$ and $Y^0$ are identical. 
	\item Positivity: $p \big( x \big| \ell \big) > 0$ for $x = 0, 1$ and for all $\ell$. Equivalently, $p \big( x, \ell \big) > 0$ for all $x, \ell$. 
\end{itemize}
We want to estimate $E \big( Y^g \big)$ for $g= 0, 1$ using observed data $O\equiv (Y, X, L)$. The g-formula approach expands this quantity as 
\begin{align}
	E \left( Y^g \right) & = \int \left\{ \int y p \left( y \big| x, \ell \right) d y \right\} p ( \ell ) d \ell\label{bgfcomments:eqn:freqestimand} 
\end{align}
by conditional exchangeability and consistency, and therefore is a function of only observed data.


\subsubsection{Parametrization}Note that we have kept the forms of probability functions unspecified thus far. Hereafter, we refer to specified forms of such probability functions as {\sl models} with corresponding {\sl parameters}. Suppose the true probabilities are parametrized such that $p \big( y \big| x, \ell; \beta \big)$ and $p ( \ell; \eta )$. Hence, we have 
\begin{align}
	E \left( Y^g; \beta, \eta \right) & = \int E \left( Y^g \big| X=g, \ell; \beta \right) p \left( \ell; \eta \right) d \ell \label{bgfcomments:eqn:paramgf1} . 
\end{align}
Expression (\ref{bgfcomments:eqn:paramgf1}) is often called the parametric g-formula in the epidemiology literature. Importantly, note the implicit modification to the following assumption. 
\begin{itemize}
	\item Parametric conditional exchangeability: $Y^g \perp\!\!\!\perp X \big| L; \beta$ for $x= 0, 1$ 
\end{itemize}
In the frequentist approach, one could use a method such as maximum likelihood estimation to estimate the fixed parameter $\beta$ using the observed values $o$, and to estimate the fixed parameter $\eta$ using the observed values $\ell$. This approach illustrates an example of a frequentist \lq\lq predictive distribution" of $Y^g$ would just be the distribution of $Y^g$ defined by substituting estimates $\hat{\beta}$ and $\hat{\eta}$ for $\beta$ and $\eta$, respectively, in $p \big( y^g; \beta, \eta \big)$ for a given value $X = g$. That is, this predictive distribution is $p \big( y^g; \hat{\beta}, \hat{\eta} \big)$. This can be seen from the fact that 
\begin{align*}
	E \left( Y^g; \beta, \eta \right)	
	& = \int y^g p \left( y^g; \beta, \eta \right) d y^g	
\end{align*}
by conditional exchangeability and consistency. In this way, the estimate $p \big( y^g; \hat{\beta}, \hat{\eta} \big)$ is a frequentist predictive distribution that can be used to calculate prediction intervals, or to calculate the estimate $E \big( Y^g; \hat{\beta}, \hat{\eta} \big)$.

\subsection{Bayesian Approach} 

\subsubsection{Bayesian G-Formula}The mean potential outcome, the g-formula estimand of interest in (\ref{bgfcomments:eqn:paramgf1}), may be estimated within the Bayesian framework as the posterior predictive mean potential outcome. The frequentist approach estimates the parametrized expected potential outcome $E \big( Y^g; \beta, \eta \big)$ by using the information in $o$ to estimate $\beta$ and $\eta$, e.g., as $\hat{\beta}$ and $\hat{\eta}$, respectively, in order to calculate $E \big( Y^g; \hat{\beta}, \hat{\eta} \big)$ as an estimate of $E \big( Y^g; \beta, \eta \big)$. Contrast this with the following Bayesian quantity, the posterior predictive distribution of the potential outcome, which explicitly conditions on $o$ and marginalizes over any parameters. 
\begin{align*}
	p \left( \tilde{y}^{g} \big| o \right) & = \int p \left( \tilde{y}^{g} \big| \theta, o \right) p \left( \theta \big| o \right) d \theta . 
\end{align*}
It can be seen that the Bayesian analogue of (\ref{bgfcomments:eqn:freqestimand}) is 
\begin{align}
	E \left( \tilde{Y}^{g} \big| o \right) & = E_\theta \left\{ E_{\tilde{Y}^{g}} \left( \tilde{Y}^{g} \big| \theta, o \right) \Big| o \right\}	
	\nonumber \\
	& = E_\theta \left[ E_{\tilde{L}} \left\{ E_{\tilde{Y}^{g}} \left( \tilde{Y}^{g} \big| \tilde{L}, \theta, o \right) \Big| \theta, o \right\} \bigg| o \right] 	
	\nonumber \\
	& = \int E_{\tilde{L}} \left\{ E_{\tilde{Y}^{g}} \left( \tilde{Y}^{g} \big| \tilde{L}, \theta, o \right) \Big| \theta, o \right\} p \left( \theta \big| o \right) d \theta 	
	\nonumber \\
	& = \int \left\{ \int E \left( \tilde{Y}^{g} \big| \tilde{\ell}, \theta, o \right) p \left( \tilde{\ell} \big| \theta, o \right) d \tilde{\ell} \right\} p \left( \theta \big| o \right) d \theta	
	\nonumber \\
	& = \int \left[ \int \left\{ \int \tilde{y}^{g} p \left( \tilde{y}^{g} \big| \tilde{\ell}, \theta, o \right) d \tilde{y}^{g} \right\} p \left( \tilde{\ell} \big| \theta, o \right) d \tilde{\ell} \right] p \left( \theta \big| o \right) d \theta 	
	\label{bgfcomments:eqn:predmean} , 
\end{align}
where $\theta$ denotes a set of parameters. It is worth noting that the component of (\ref{bgfcomments:eqn:predmean}) that resembles the parametric g-formula (\ref{bgfcomments:eqn:paramgf1}) is 
\begin{equation}
	\int E \left( \tilde{Y}^{g} \big| \tilde{\ell}, \theta, o \right) p \left( \tilde{\ell} \big| \theta, o \right) d \tilde{\ell} \label{bgfcomments:eqn:bgfunmarginalized} , 
\end{equation}
which is equal to \( E \left( \tilde{Y}^{g} \big| \theta, o \right) \) by the equivalence of lines 1 and 4 of (\ref{bgfcomments:eqn:predmean}). In the following subsections, we will apply the assumptions in subsection \ref{bgfcomments:subsec:assumptions}, which will be shown to imply that the quantity \( E \left( \tilde{Y}^{g} \big| \tilde{\ell}, \theta, o \right) \) in (\ref{bgfcomments:eqn:bgfunmarginalized}) can be identified using only observed or observable data.

\subsubsection{Assumptions}\label{bgfcomments:subsec:assumptions}In all following subsections, we assume the following. We do not include the positivity assumption here, as it will first be motivated through the derivations, and then included in the summary in subsection \ref{bgfcomments:subsec:summary}. Let $\theta = \big\{ \beta, \alpha, \eta \big\}$. 
\begin{description}
	\item [Bayesian conditional exchangeability] $Y^g \perp\!\!\!\perp X \big| L, \theta$ for $g= 0, 1$ 
	\item [Consistency] $Y = Y^1 X + Y^0 ( 1 - X )$ where the support of $Y^1$ and $Y^0$ are identical. 
	\item []We also make the following set of independence assumptions regarding the parameters
	\begin{description}
		\item[I1] $\big\{ \tilde{A} \big\} \perp\!\!\!\perp O \big| \theta$ for any set $\big\{ \tilde{A} \big\}$ that is a subset of $\big\{ \tilde{Y}, g, \tilde{L} \big\}$. Conditional on $\theta$, any subset of future outcomes $\big\{ \tilde{A} \big\}$ of $\big\{ \tilde{Y}, g, \tilde{L} \big\}$ is independent of all observed outcomes $O$. 
		\item [I2] $\beta$, $\alpha$, and $\eta$ are mutually and jointly independent 
		\item [I3] $Y \perp\!\!\!\perp \alpha, \eta \big| X, L, \beta \Leftrightarrow p \big( y \big| x, \ell, \theta \big) = p \big( y \big| x, \ell, \beta \big)$ 
		\item [I4] $X, L \perp\!\!\!\perp \beta \big| \alpha, \eta \Leftrightarrow p \big( x, \ell \big| \theta \big) = p \big( x, \ell \big| \alpha, \eta \big)$ 
		\item [I5] $X \perp\!\!\!\perp \eta \big| L, \alpha \Leftrightarrow p \big( x \big| \ell, \alpha, \eta \big) = p \big( x \big| \ell, \alpha \big)$ 
		\item [I6] $L \perp\!\!\!\perp \beta \big| \alpha, \eta \Leftrightarrow p \big( \ell \big| \theta \big) = p \big( \ell \big| \alpha, \eta \big)$ and $L \perp\!\!\!\perp \alpha \big| \eta \Leftrightarrow p \big( \ell \big| \alpha, \eta \big) = p \big( \ell \big| \eta \big)$ 
	\end{description}
\end{description}

\subsubsection{Conditional Distribution of $\tilde{y}$}The conditional distribution of $\tilde{y}^{g}$ in (\ref{bgfcomments:eqn:predmean}) is 
\begin{align*}
	p \left( \tilde{y}^{g} \big| \tilde{\ell}, \theta, o \right) & = p \left( \tilde{y}^{g} \big| \tilde{\ell}, y, x, \ell, \theta \right) = \frac { p \left( \tilde{y}^{g}, \tilde{\ell}, y, x, \ell, \theta \right) } { p \left( \tilde{\ell}, y, x, \ell, \theta \right) } 	
	. 
\end{align*}
The probability function in the numerator is 
\begin{align}
	& p \left( \tilde{y}^{g}, \tilde{\ell}, y, x, \ell, \theta \right) \nonumber \\
	& = \left\{ p \left( \tilde{y} \big| g, \tilde{\ell}, \beta \right) p \left( y \big| x, \ell, \beta \right) \pi \left( \beta \right) \right\} \left\{ p \left( \tilde{\ell} \big| \eta \right) p \left( x, \ell \big| \alpha, \eta \right) \pi \left( \alpha \right) \pi \left( \eta \right) \right\} \label{bgfcomments:eqn:numeratorY}	
\end{align}
by Bayesian conditional exchangeability, consistency, I1-I4, and I6. Likewise, the probability function in the denominator is 
\begin{align}
	p \left( \tilde{\ell}, y, x, \ell, \theta \right) & = \left\{ p \left( y \big| x, \ell, \beta \right) \pi \left( \beta \right) \right\} \left\{ p \left( \tilde{\ell} \big| \eta \right) p \left( x, \ell \big| \alpha, \eta \right) \pi \left( \alpha \right) \pi \left( \eta \right) \right\} \label{bgfcomments:eqn:denominatorY}	
\end{align}
by I1-I4 and I6. Hence, {\small 
\begin{align}
	p \left( \tilde{y} \big| g, \tilde{\ell}, \theta, o \right) & = \frac { \left\{ p \left( \tilde{y} \big| g, \tilde{\ell}, \beta \right) p \left( y \big| x, \ell, \beta \right) \pi \left( \beta \right) \right\} \left\{ p \left( \tilde{\ell} \big| \eta \right) p \left( x, \ell \big| \alpha, \eta \right) \pi \left( \alpha \right) \pi \left( \eta \right) \right\} } { \left\{ p \left( y \big| x, \ell, \beta \right) \pi \left( \beta \right) \right\} \left\{ p \left( \tilde{\ell} \big| \eta \right) p \left( x, \ell \big| \alpha, \eta \right) \pi \left( \alpha \right) \pi \left( \eta \right) \right\} } = p \left( \tilde{y} \big| g, \tilde{\ell}, \beta \right) \label{bgfcomments:eqn:predmeanexpectY} . 
\end{align}
} 

\subsubsection{Conditional Distribution of $\tilde{\ell}$}The conditional distribution of $\tilde{\ell}$ in (\ref{bgfcomments:eqn:predmean}) is 
\begin{align}
	p \left( \tilde{\ell} \big| \theta, o \right) & = p \left( \tilde{\ell} \big| \eta \right) \label{bgfcomments:eqn:predmeanpdfL}	
\end{align}
by I1, I2, and I6.

\subsubsection{Posterior Distributions}The posterior distribution of $\theta$ can be expanded as 
\begin{align}
	p \left( \theta \big| o \right) & = \frac { p \left( o \big| \theta \right) \pi \left( \theta \right) } { \int p \left( o \big| \theta \right) \pi \left( \theta \right) d \theta }	
	\label{bgfcomments:eqn:posteriorthetaunexpanded} . 
\end{align}
The conditional probability function in the numerator and denominator is 
\begin{align}
	p \left( o \big| \theta \right) & = p \left( y \big| x, \ell, \beta \right) p \left( x \big| \ell, \alpha \right) p \left( \ell \big| \eta \right)	%
	\label{bgfcomments:eqn:likelihood}	
\end{align}
by I3-I6. Hence, 
\begin{align}
	p \left( \theta \big| o \right) & = \left\{ \frac { p \left( y \big| x, \ell, \beta \right) \pi \left( \beta \right) } { \int p \left( y \big| x, \ell, \beta \right) \pi \left( \beta \right) d \beta } \right\} \left\{ \frac { p \left( x \big| \ell, \alpha \right) \pi \left( \alpha \right) } { \int p \left( x \big| \ell, \alpha \right) \pi \left( \alpha \right) d \alpha } \right\} \left\{ \frac { p \left( \ell \big| \eta \right) \pi \left( \eta \right) } { \int p \left( \ell \big| \eta \right) \pi \left( \eta \right) d \eta } \right\} \label{bgfcomments:eqn:posteriortheta}	
\end{align}
by I2. Note that the posterior distribution of $\beta$ is 
\begin{align}
	p \left( \beta \big| y, x, \ell \right) & = \frac { p \left( y \big| x, \ell, \beta \right) \pi \left( \beta \right) } { \int p \left( y \big| x, \ell, \beta \right) \pi \left( \beta \right) d \beta } \label{bgfcomments:eqn:posteriorbeta}	
\end{align}
by I2-I4. Also note that the posterior distribution of $\alpha$ is 
\begin{align*}
	p \left( \alpha \big| x, \ell \right) & = \frac { p \left( x \big| \ell, \alpha \right) \pi \left( \alpha \right) } { \int p \left( x \big| \ell, \alpha \right) \pi \left( \alpha \right) d \alpha }	
\end{align*}
by I2, I5, and I6. We therefore have 
\begin{align}
	p \left( \theta \big| o \right) & = p \left( \beta \big| y, x, \ell \right) p \left( \alpha \big| x, \ell \right) p \left( \eta \big| \ell \right) \label{bgfcomments:eqn:postdistns} . 
\end{align}

\subsubsection{Two Equalities}Consider the following equality, whereby the Bayesian analogue of (\ref{bgfcomments:eqn:freqestimand}) is instead expanded as 
\begin{align}
	E \left( \tilde{Y}^{g} \big| o \right)
	& = \int E \left( \tilde{Y}^{g} \big| \tilde{\ell}, o \right) p \left( \tilde{\ell} \big| o \right) d \tilde{\ell}
	\label{bgfcomments:eqn:altpredmean} . 
\end{align}
The posterior predictive distribution of $\tilde{\ell}$ is 
\begin{align*}
	p \left( \tilde{\ell} \big| o \right) & = \frac { \int p \left( \tilde{\ell}, \ell, y, x, \theta \right) d \theta } { \int p \left( \ell, y, x, \theta \right) d \theta }
	. 
\end{align*}
The probability function in the numerator is 
\begin{align*}	
	p \left( \tilde{\ell}, y, x, \ell, \theta \right) & = \left\{ p \left( y \big| x, \ell, \beta \right) \pi \left( \beta \right) \right\} \left\{ p \left( x \big| \ell, \alpha \right) \pi \left( \alpha \right) \right\} \left\{ p \left( \tilde{\ell} \big| \eta \right) p \left( \ell \big| \eta \right) \pi \left( \eta \right) \right\}	%
\end{align*}
by I1-I6. Likewise, the probability function in the denominator is 
\begin{align*}
	p \left( y, x, \ell, \theta \right) & = \left\{ p \left( y \big| x, \ell, \beta \right) \pi \left( \beta \right) \right\} \left\{ p \left( x \big| \ell, \alpha \right) \pi \left( \alpha \right) \right\} \left\{ p \left( \ell \big| \eta \right) \pi \left( \eta \right) \right\}
\end{align*}
by I2-I6. Hence, the equality 
\begin{align}
	p \left( \tilde{\ell} \big| o \right) & = \int p \left( \tilde{\ell} \big| \eta \right) p \left( \eta \big| \ell \right) d \eta = p \left( \tilde{\ell} \big| \ell \right) \label{bgfcomments:eqn:altpredmeanpdfL} . 
\end{align}

\subsubsection{Summary of Calculation Requirements}\label{bgfcomments:subsec:summary}The relevant parameter posterior distributions are 
\begin{align*}
	p \left( \beta \big| y, x, \ell \right) & = \frac { \mathcal{L} \left( \beta \big| x, \ell, y \right) \pi \left( \beta \right) } { \int \mathcal{L} \left( \beta \big| x, \ell, y \right) \pi \left( \beta \right) d \beta } 
\end{align*}
by (\ref{bgfcomments:eqn:posteriorbeta}), and 
\begin{align*}
	p \left( \eta \big| \ell \right) & = \frac { \mathcal{L} \left( \eta \big| \ell \right) \pi \left( \eta \right) } { \int \mathcal{L} \left( \eta \big| \ell \right) \pi \left( \eta \right) d \eta } , 
\end{align*}
where $\mathcal{L} ( \cdot )$ denotes the likelihood function. Let $O_i = \big\{ Y_i, X_i, L_i \big\}$ represent the observed data for measurement $i$. For example, for an observed sample of $i = 1, \hdots, n$ independent measurements, we have 
\begin{align*}
	\mathcal{L} \left( \beta \big| x, \ell, y \right) & = \prod_{i=1}^n p \left( y_i \big| x_i, \ell_i, \beta \right) , \\
	\mathcal{L} \left( \eta \big| \ell \right) & = \prod_{i=1}^n p \left( \ell_i \big| \eta \right) . 
\end{align*}
Expressions (\ref{bgfcomments:eqn:predmeanexpectY}), (\ref{bgfcomments:eqn:predmeanpdfL}), and (\ref{bgfcomments:eqn:postdistns}) may now be substituted into expression (\ref{bgfcomments:eqn:predmean}) for the posterior predictive mean, yielding 
\begin{align}
	E \left( \tilde{Y}^{g} \big| o \right) & = \int \left[ \int \left\{ \int_{ \tilde{y}^{g} } \tilde{y} p \left( \tilde{y} \big| g, \tilde{\ell}, \theta, o \right) d \tilde{y} \right\} p \left( \tilde{\ell} \big| \theta, o \right) d \tilde{\ell} \right] p \left( \theta \big| o \right) d \theta \nonumber \\
	& = \int \int \int \left[ \int \left\{ \int_{ \tilde{y}^{g} } \tilde{y} p \left( \tilde{y} \big| g, \tilde{\ell}, \beta \right) d \tilde{y} \right\} p \left( \tilde{\ell} \big| \eta \right) d \tilde{\ell} \right] p \left( \beta \big| y, x, \ell \right) p \left( \alpha \big| x, \ell \right) p \left( \eta \big| \ell \right) d \eta d \alpha d \beta \label{bgfcomments:eqn:predmeanexpanded} . 
\end{align}
Note that (\ref{bgfcomments:eqn:predmeanexpanded}) is well defined only if all denominator quantities used in its definition are not equal to zero. That is, the following additional assumptions have been made implicitly up to this point. 
\begin{itemize}
	\item The expanded denominator in (\ref{bgfcomments:eqn:denominatorY}) is greater than 0 if $p \big( \tilde{\ell} \big| \eta \big) > 0$, $p \big( y \big| x, \ell, \beta \big) > 0$, $p \big( x \big| \ell, \alpha \big) > 0$ by I5 and $p \big( \ell \big| \eta \big) > 0$ by I6, and $\pi \big( \theta \big) > 0$. Note that $p \big( \tilde{\ell} \big| \eta \big) > 0$ and $p \big( \ell \big| \eta \big) > 0$ are implied by assuming $p \big( \ell \big| \eta \big) > 0$ for all $\ell$. 
	\item The denominator used to define (\ref{bgfcomments:eqn:predmeanpdfL}) is greater than 0 if $\pi \big( \theta \big) > 0$. 
	\item The denominator in $(\ref{bgfcomments:eqn:posteriorthetaunexpanded})$ is greater than 0 if $p \big( o \big) > 0$. 
\end{itemize}
For the support (i.e., allowed values) of the random variable $\theta$, $p \big( \theta \big) > 0$ is always true. Note that $p \big( o \big) > 0$ by definition because $o$ represents observed quantities. Hence, the key assumption that must me made is: 
\begin{description}
	\item [Bayesian positivity] $p \big( \ell \big| \eta \big) > 0$ for all $\ell$ and $\eta$; $p \big( y \big| x, \ell, \beta \big) > 0$ for all $y$, $x$, $\ell$, and $\beta$; and $p \big( x \big| \ell, \alpha \big) > 0$ for all $x$, $\ell$, and $\alpha$. 
\end{description}
One reason to explicitly acknowledge the Bayesian positivity assumption is that it may help set realistic constraints on the possible values of $\theta = \big\{ \beta, \alpha, \eta \big\}$. The following variable values and models must be observed, set, or specified. 
\begin{itemize}
	\item Observed values: $y, x, \ell$ 
	\item Values set by the intervention: $g$ 
	\item Models specified: \( p \left( y \big| x, \ell, \beta \right) , p \left( \ell \big| \eta \right) , \pi \left( \beta \right) , \pi \left( \eta \right) \) 
\end{itemize}
Importantly, note that $\tilde{\ell}$ need not be set because it is integrated out.

\subsubsection{Some Binary Outcome Results}Suppose the outcome is binary such that $\tilde{Y}^{g} = 0, 1$ for $x = 0, 1$. Hence,
\[ E \left( \tilde{Y}^{g} \big| z \right) = \Pr \left( \tilde{Y}^{g} = 1 \big| z \right) \]
by the definition of an expectation for any set of variable realizations $z$ (including the empty set). Suppose assumptions Bayesian conditional exchangeability, consistency, and I1-I6 are true. From the equivalence of lines 1 and 2 of (\ref{bgfcomments:eqn:predmeanexpanded}), \( E \left( \tilde{Y}^{g} \big| \theta, o \right) = E \left( \tilde{Y}^{g} \big| \theta \right) . \) Hence, (\ref{bgfcomments:eqn:predmean}) is therefore equal to 
\begin{align*}
	\Pr \left( \tilde{Y}^{g} = 1 \big| o \right) & = \int \Pr \left( \tilde{Y}^{g} = 1 \big| \theta \right) p \left( \theta \big| o \right) d \theta	
	. 
\end{align*}
Equivalently, 
\begin{align}
	p \left( \tilde{y}^{g} \big| o \right) & = \int p \left( \tilde{y}^{g} \big| \theta \right) p \left( \theta \big| o \right) d \theta \label{bgfcomments:eqn:predmeanbinaryoutcome} . 
\end{align}
Similarly, (\ref{bgfcomments:eqn:altpredmean}) is equal to 
\begin{align*}
	\Pr \left( \tilde{Y}^{g} = 1 \big| o \right) & = \int \Pr \left( \tilde{Y}^{g} = 1 \big| \tilde{\ell}, o \right) p \left( \tilde{\ell} \big| o \right) d \tilde{\ell} 
\end{align*}
Equivalently, 
\begin{align}
	p \left( \tilde{y}^{g} \big| o \right) & = \int p \left( \tilde{y} \big| g, \tilde{\ell}, o \right) p \left( \tilde{\ell} \big| o \right) d \tilde{\ell}	
	\label{bgfcomments:eqn:altpredmeanbinaryoutcome}	
\end{align}
by conditional exchangeability and consistency. Finally, from the equivalence of lines 1 and 2 of (\ref{bgfcomments:eqn:predmeanexpanded}) and the penultimate line of (\ref{bgfcomments:eqn:predmean}), 
\begin{align*}
	\Pr \left( \tilde{Y}^{g} = 1 \big| o \right) & = \int \left\{ \int \Pr \left( \tilde{Y} = 1 \big| g, \tilde{\ell}, \theta \right) p \left( \tilde{\ell} \big| \theta \right) d \tilde{\ell} \right\} p \left( \theta \big| o \right) d \theta 
\end{align*}
by conditional exchangeability and consistency. Equivalently, 
\begin{align}
	p \left( \tilde{y}^{g} \big| o \right) & = \int \int p \left( \tilde{y} \big| g, \tilde{\ell}, \theta \right) p \left( \tilde{\ell} \big| \theta \right) p \left( \theta \big| o \right) d \theta d \tilde{\ell} \label{bgfcomments:eqn:ppdposhort} 
\end{align}
and 
\begin{align}
	p \left( \tilde{y}^{g} \big| o \right) & \propto \int \int \int p \left( \tilde{y} \big| g, \tilde{\ell}, \beta \right) p \left( \tilde{\ell} \big| \eta \right) \mathcal{L} \left( \beta \big| y, x, \ell \right) \pi \left( \beta \right) \mathcal{L} \left( \eta \big| \ell \right) \pi \left( \eta \right) d \beta d \eta d \tilde{\ell} \label{bgfcomments:eqn:ppdpofull}	
\end{align}
by (\ref{bgfcomments:eqn:postdistns}), I3, and I6. Suppose $Y^g$ (and therefore, $Y$ by consistency), $X$, and $L$ are binary such that $y^g = 0, 1$, $x = 0, 1$, and $\ell = 0, 1$. For an observed sample of $i = 1, \hdots, n$ independent measurements, we have 
\begin{align}
	\mathcal{L} \left( \theta \big| o \right) & = \prod_{i=1}^n \Pr \left( Y_i = 1 \big| X_i = x_i, L_i = \ell_i, \beta \right)^{y_i} \Pr \left( Y_i = 0 \big| X_i = x_i, L_i = \ell_i, \beta \right)^{1 - y_i} \times \nonumber \\
	& \quad \quad \; \; \Pr \left( X_i = 1 \big| L_i = \ell_i, \alpha \right)^{x_i} \Pr \left( X_i = 0 \big| L_i = \ell_i, \alpha \right)^{1 - x_i} \times \nonumber \\
	& \quad \quad \; \; \Pr \left( L_i = 1 \big| \eta \right)^{\ell_i} \Pr \left( L_i = 0 \big| \eta \right)^{1 - \ell_i} \label{bgfcomments:eqn:likelihoodbinary}	
\end{align}
by (\ref{bgfcomments:eqn:likelihood}). Note that 
\begin{align}
	\mathcal{L} \left( \theta \big| o \right) & = \mathcal{L} \left( \beta \big| y, x, \ell \right) \mathcal{L} \left( \alpha \big| x, \ell \right) \mathcal{L} \left( \eta \big| \ell \right) \label{bgfcomments:eqn:likelihoodfactorization}	
\end{align}
by (\ref{bgfcomments:eqn:likelihood}).
\subsection{Time-varying quantities} We have shown that the posterior predictive distribution that we would have observed under some treatment regime $g$ is given by the Bayesian g-formula in the simple, time-fixed setting. As noted in the manuscript, we are often also interested in the causal effects of exposure regimes that may vary over time and/or be subject to confounding by other time-varying factors. Following, Theorem 3.1 by Robins (1997) \cite{robins1997causal}, the generalization of our results to this time varying setting requires little more than what has already been shown. The proof of this theorem generalizes our results by considering the the time fixed setting to be equivalent to a "one step ahead innovation" where the counterfactual world at time $t$ under regime $g$ is linked to the observed data at time $t$ among individuals who have been observed to follow $g$ (within strata of prior exposure and covariates). This theorem states that, given that we have had some individuals who the distribution of the counterfactual outcome $Y^g(t+1)$, given $L(t)$ is identified among those whose previous exposure or treatment is consistent with $g.$ Thus, all that is required to generalize our results to the time-varying setting is to modify Bayesian conditional exchangeability, consistency, and Bayesian positivity to consider the entire histories of $\bar{L}(t)\equiv(L(0), ..., L(t))$ and $\bar{X}(t)\equiv(X(0), ..., X(t))$, rather than simply the vector $(L,X)$  as time-fixed quantities.

\end{document}